\documentclass[sigconf,nonacm]{acmart}

\renewcommand\footnotetextcopyrightpermission[1]{}
\usepackage{booktabs}
\usepackage{amsmath}
\usepackage{url}
\usepackage[htt]{hyphenat}
\begin{document}

\title[GPUSparse: Exact Learned Sparse Retrieval on a GPU]{GPUSparse: Fast and
Exact Learned Sparse Retrieval\\on a GPU-Resident Inverted Index}

\author{Ashutosh Sharma}
\affiliation{%
  \institution{MIT-IBM Watson AI Lab}
  \country{USA}
}
\email{ashutosh.sharma7@ibm.com}

\begin{abstract}
Learned sparse models such as SPLADE match the quality of dense bi-encoders and
keep the exact-match behavior of lexical retrieval. They are served on CPUs, using
inverted-index traversal algorithms such as WAND. These algorithms are fast because
they prune: they skip documents that cannot enter the top-$k$. Pruning makes the
result approximate, and its control flow is sequential and data-dependent, which
suits a CPU and not a GPU.

We argue that a GPU should not run a faster traversal. It should remove traversal
altogether. A GPU has enough memory bandwidth to score every posting, so exactness
need not be traded for speed. We present \textsc{GPUSparse}, which keeps the
inverted index in GPU memory and scores queries with one fused scatter-add kernel.
Nothing is pruned.

\textsc{GPUSparse} searches all 8.8M MS~MARCO passages in 1.27~ms per query, at 787
queries per second on one H100. Its results are exact: Recall@10, @100 and @1000
against an exhaustive FP32 reference are all 1.000, and quality at full scale
matches Pyserini's exact SPLADE retrieval to three decimals.

Exactness costs hardware rather than latency. Under core pinning, one H100 is worth
about seven CPU cores. \textsc{GPUSparse} is $5.6\times$ faster than one core and
$2.8\times$ slower than a 32-core socket. A CPU socket is therefore cheaper for
approximate serving. The GPU is the better choice when exactness is required, where
\textsc{GPUSparse} is $3.0\times$ faster than exact CPU block-max pruning. We also
decompose GPU latency. Traversal takes 55 to 80\% of it and selection 18 to 38\%,
and the split moves with scale, which reconciles three conflicting published
reports.
\end{abstract}

\maketitle

\renewcommand{\thefootnote}{\fnsymbol{footnote}}
\footnotetext[1]{Code: \url{https://github.com/ashutoshuiuc/gpu-sparse}.}
\renewcommand{\thefootnote}{\arabic{footnote}}

\section{Introduction}
\label{sec:intro}

Learned sparse models map text to sparse vectors over a vocabulary. SPLADE and its
successors~\cite{formal2021splade,formal2021spladev2,formal2022distillation,lassance2024splade}
match the quality of dense bi-encoders~\cite{karpukhin2020dense}. They also keep the
exact-match behavior of lexical retrieval~\cite{robertson2009probabilistic}. These
models are served on CPUs, using inverted-index traversal algorithms such as
WAND~\cite{broder2003efficient} and block-max WAND~\cite{ding2011faster}. Dense
retrieval moved to GPUs long ago, because scoring $N$ documents is one matrix
product.

Two properties of CPU traversal have kept learned sparse retrieval on the CPU.
First, these algorithms are fast because they prune. They keep an upper bound on
what each posting list can still contribute, and skip documents that cannot enter
the top-$k$. Pruning is exact only if the bounds are true bounds. The fastest
current systems~\cite{bruch2024seismic,mallia2024bmp} relax that condition to gain
latency, so their results are approximate. Second, the pruning logic is sequential and
data-dependent. Each skip depends on the current $k$-th best score, and a GPU runs
threads in groups of 32 that must follow the same branch. Data-dependent skipping
makes those groups diverge.

In this paper we argue that a GPU should not run a faster traversal. It should
remove traversal altogether. Scoring every posting does more arithmetic than
pruning does, but a GPU has bandwidth to spare, so the extra work is cheap. Once
the work is cheap, exactness need not be traded for speed.

We present \textsc{GPUSparse}. It keeps the inverted index in GPU memory as two
flat arrays, and scores queries with one fused Triton kernel. Each program reads
one query term's posting list and adds its weighted contributions into a per-query
score buffer, using atomic adds. No document is skipped and no score is
approximated.

\textsc{GPUSparse} searches all 8.8M MS~MARCO passages in 1.27~ms per query, at 787
queries per second on one H100. The index takes 7.91~GiB, about 10\% of the device.
The results are exact. Against an exhaustive FP32 reference, Recall@10, @100 and
@1000 are all exactly 1.000 at 100K, 500K and 1M documents. At full scale MRR@10,
nDCG@10 and Recall@1000 match Pyserini's exact SPLADE retrieval to three decimals.

We then ask what the design costs, and we answer with measured baselines rather
than estimates. Normalized by hardware, one H100 is worth about seven pinned CPU
cores. Ding et al.~\cite{ding2009gpu} reached the same conclusion in 2009 from a
GPU inverted index, writing that ``currently a quad-core CPU still provides better
value than a GPU''. The conclusion has now survived seventeen years and three GPU
generations. We read that as evidence about the workload rather than about either
generation of hardware. The work per posting is small and irregular, and a GPU must
spend bandwidth on a dense accumulator to expose parallelism. What we add is the
point at which the conclusion reverses. When exactness is required at $k{=}1000$,
\textsc{GPUSparse} is $3.0\times$ faster than exact CPU pruning. When a 3\% recall
loss is acceptable at $k{=}10$, a CPU socket is both faster and cheaper.

Finally, we measure where GPU time goes, because that determines what to optimize
next. Traversal takes 55 to 80\% of it, and its share grows with the collection.
Selection takes 18 to 38\%, and its share shrinks. Published reports disagree here:
FAERY reports up to 80.4\% for selection~\cite{zeng2022faery}, AgentIR 75 to
95\%~\cite{yuan2026agentir}, and FAISS 20.6\% at large
scale~\cite{johnson2021billion}. Our sweep is consistent with all three once scale
and batch size are fixed.

We summarize our contributions as follows.

\begin{itemize}
\item \textbf{Exact retrieval at full scale.} Recall@10, @100 and @1000 are all
exactly 1.000 against an exhaustive reference, at 1.27~ms per query on all 8.8M
passages (\S\ref{sec:quality}). The closest prior GPU system excludes this
collection, because its score buffer exceeds a 16~GB cap.

\item \textbf{A hardware-normalized cost comparison.} One H100 is worth about seven
pinned CPU cores (\S\ref{sec:cpu}). That places the GPU behind a CPU socket for
approximate serving, and $3.0\times$ ahead of it when exactness is required.

\item \textbf{A latency decomposition.} The traversal and selection shares depend
on collection size and batch size (\S\ref{sec:breakdown}). This reconciles three
conflicting published reports. A second profiler agrees with ours to within 0.3
percentage points.

\item \textbf{Two measurement traps.} \texttt{pyseismic-lsr} ignores its
\texttt{num\_threads} argument and uses 34 to 36 cores by default
(\S\ref{sec:cpu}). An unpinned run therefore reads as a $3\times$ CPU advantage
when it compares a socket with one accelerator. Separately,
percent-of-peak bandwidth from an analytic byte count is wrong here by $42\times$
and $1.56\times$, in opposite directions (\S\ref{sec:bandwidth}).

\item \textbf{A negative result on fused selection.} Threshold compaction cuts
selector input by $1{,}873\times$ and gains only $1.13\times$ end to end
(\S\ref{sec:fusedtopk}). The remaining headroom is in materializing the score
buffer, not in the selection algorithm.
\end{itemize}

\section{Background}
\label{sec:background}

We first describe how a GPU spends time, because that is what decides whether
scoring every posting is affordable. We then describe learned sparse retrieval and
how it is served today. Related work is discussed in \S\ref{sec:related}.

\subsection{GPU Performance Characteristics}
\label{sec:gpuperf}

\textbf{Memory hierarchy.} A GPU has a few levels of memory, and the smaller
levels are faster. An H100 SXM 80GB has 3.35~TB/s of HBM3 bandwidth, a 50~MB L2
cache, and up to 228~KB of shared memory on each of its 132 streaming
multiprocessors.
On-chip memory is far faster than HBM and far smaller.

\textbf{Execution model.} A GPU runs one function, called a kernel, across many
threads at once. Threads are grouped into warps of 32 that issue the same
instruction. A kernel reads its inputs from HBM, computes, and writes its outputs
back to HBM. Two properties follow. A warp is fastest when its 32 threads read
consecutive addresses, which the hardware coalesces into one transaction. A warp is
slowest when its threads take different branches, because the hardware then runs
the branches one after another.

\textbf{Memory-bound and compute-bound work.} An operation is compute-bound if its
arithmetic takes longer than its memory accesses, and memory-bound otherwise. The
usual indicator is arithmetic intensity, the number of arithmetic operations per
byte read~\cite{williams2009roofline}. Retrieval scoring is memory-bound. Each
posting contributes one multiply and one add for eight bytes read, so the kernel
spends its time moving data. This is what makes exact scoring affordable: the extra
arithmetic is close to free, and only the extra bytes cost time.

\textbf{Kernel fusion.} Memory-bound work is usually accelerated by fusing kernels.
If several operations read the same input, a fused kernel reads it once instead of
once per operation. Fusion also avoids writing intermediates to HBM. Both matter
here, as \S\ref{sec:scoring} shows.

\subsection{Learned Sparse Retrieval}
\label{sec:lsr}

A model maps text $x$ to $\mathbf{s}(x) \in \mathbb{R}_{\geq 0}^{|\mathcal{V}|}$
with few non-zeros. SPLADE v2~\cite{formal2021spladev2} max-pools over tokens,
\begin{equation}
    \mathbf{s}(x) = \max_{t \in x} \log(1 + \text{ReLU}(\mathbf{W} \mathbf{h}_t + \mathbf{b})),
\end{equation}
where original SPLADE~\cite{formal2021splade} sum-pools the same per-token term.
A score is an inner product, $\text{score}(q,d) = \mathbf{s}(q)^\top
\mathbf{s}(d)$. On MS~MARCO a document has about 120 non-zero terms and a query
about 45, so most of the $30{,}522$-dimensional vocabulary is absent from any one
vector.

\subsection{How It Is Served Today}
\label{sec:standard}

The standard implementation stores an inverted index and traverses it with
pruning. WAND~\cite{broder2003efficient} keeps one iterator per query term, sorted
by document id, and skips documents that cannot enter the top-$k$. It is exact
given true upper bounds. Block-max WAND~\cite{ding2011faster} adds per-block
maxima. Block-max pruning~\cite{mallia2024bmp} adapts this to learned sparse
vectors, and dynamic superblock pruning~\cite{carlson2025dynamic} extends it.
Seismic~\cite{bruch2024seismic} takes a different route. It clusters each posting
list into geometric blocks with summary vectors, then skips whole blocks. Two
parameters control it, a query-term cut and a block-skipping factor, and both make
it approximate.

Seismic is the strongest CPU baseline we compare against, so its published
operating point matters. Its reported single-threaded latency is 187 to
531~$\mu$s at $k{=}10$, at 90 to 97\% accuracy. Independent studies report
Recall@10 of 0.977~\cite{song2026static} and Recall@50 above
99\%~\cite{li2026sindi}. Configured inside its documented range, Seismic is
therefore close to exact, and \S\ref{sec:cpu} reproduces that result.

Two things in this design do not transfer to a GPU. The skipping decisions are
sequential, since each one depends on the current $k$-th best score, and they are
data-dependent, so neighbouring threads would take different branches. A GPU
implementation of pruning therefore gives up the parallelism that makes the GPU
fast. \S\ref{sec:design} takes the opposite approach and prunes nothing.

\section{GPUSparse}
\label{sec:design}

We now describe the design. \S\ref{sec:index} gives the index layout,
\S\ref{sec:scoring} the fused scoring kernel, and \S\ref{sec:chunked} a chunked
variant that makes peak memory independent of collection size. Everything follows
from one decision: we score every posting and prune nothing, trading work
efficiency for parallelism.

\subsection{GPU-Resident Inverted Index}
\label{sec:index}

The layout exists to make every read coalesced. We store the index as two flat
arrays, \texttt{doc\_ids} (int32) and
\texttt{scores} (float32), with per-term \texttt{offsets},
\texttt{lengths} and \texttt{padded\_lengths}. Each posting list is padded to a
multiple of the warp size $W{=}32$, so a warp loads a full chunk without
masking:
\begin{equation}
    \text{padded\_length}(t) = \lceil |\text{PL}(t)| / W \rceil \times W .
\end{equation}
Padding entries use \texttt{doc\_id}${}={-1}$ and \texttt{score}${}={0}$ and are
masked during scoring. Document ids are stored uncompressed, with no delta
coding, to avoid sequential decompression during traversal. The layout is a
variant of the 128-element blocked layouts of Ding et al.\ and Mallia et al.,
with the block width set to the warp size.

Posting lists must be ascending in document id, because the chunked path of
\S\ref{sec:chunked} locates per-chunk sub-ranges by binary search. A
non-ascending list produces incorrect sub-ranges and therefore incorrect scores
without raising an error, so the builder validates the ordering instead of
assuming it. We verify the invariant directly: across the 26,972 terms at 100K,
28,023 at 500K and 28,245 at 1M that hold two or more postings, no list is
non-ascending. The check is necessary for any builder using this layout, because
an unstable sort does not guarantee the ordering.

Index size is $N \bar{k} (4+4)(1+\epsilon_{\text{pad}})$ bytes for $N$
documents with $\bar{k}$ non-zero terms each. At $\bar{k}{=}120.5$ measured on
MS~MARCO, 100K documents need 93~MB and the full 8.8M collection needs
7.91~GiB, about 10\% of an 80~GB H100.

\subsection{Fused Scatter-Add Scoring}
\label{sec:scoring}

For query $q$ with terms $\{(t_i,w_i)\}$, the score of document $d$ is
$\sum_i w_i\, s_d(t_i)$, which decomposes into $|q|$ independent scatter-adds:
\begin{equation}
    \texttt{scores}[d] \mathrel{+}= w_i \cdot \texttt{PL}(t_i)[d], \quad \forall d \in \texttt{PL}(t_i).
\end{equation}
For $B$ queries with at most $M$ terms each we launch a $B \times M$ grid. Each
program loads one posting list and accumulates into a $[B,N]$ buffer with
\texttt{tl.atomic\_add}, processing the list in chunks of
\texttt{BLOCK\_PL} elements. Measured over
$\texttt{BLOCK\_PL} \in \{32,64,128,256,512,1024\}$ at 100K documents with
$B{=}500$, every value returns an identical top-10, and latency spans 15.8\%
across the full range but only \textbf{6.8\%} from 64 to 512 (3.53, 3.34, 3.31,
3.30~ms). The fastest is 512 at 3.302~ms. We run 128, which is 1.2\% slower, so
this parameter has little effect within that range.

Fusion is necessary for the reformulation to be practical. Written out in PyTorch as one
\texttt{index\_add\_} call per query term, the same algorithm issues roughly 64,000
kernel launches per 500-query batch. Launch overhead then sets its latency, not memory
traffic. Its latency stays flat as the collection grows, which is what launch-bound
behavior looks like.

\paragraph{Scope of the exactness claim.}
Every posting is read and every score is a complete inner product, so no document
is skipped and no score is approximated. Accumulation through
\texttt{atomic\_add} still leaves the summation order dependent on how
blocks interleave, which varies between runs, so two executions on identical
inputs are not bitwise equal. We quantify the effect: scores differ by at most
$3.8\times10^{-6}$, top-10 and top-1000 agreement are both exactly 1.000, and
about one document per query lies within $10^{-5}$ of the $k$-th best
score. The claim is therefore that the returned set is correct, not that the
final position is reproducible bit for bit. Bitwise determinism would require a
fixed reduction order, which would remove the parallelism the design depends
on.

\subsection{Chunked Scoring}
\label{sec:chunked}

The $[B,N]$ accumulator is the only part of the design whose cost grows with batch
size times collection size. At 8.84M documents it needs 33.8~MiB per query, however
few terms that query has. It is therefore what limits batch size, not the 7.91~GiB
index.

We partition documents into chunks of $C$. For each chunk we scatter-add only the
postings whose document id falls inside it, select that chunk's top-$k$, and merge
into a running top-$k$. Peak score memory becomes $O(BC + Bk)$ instead of $O(BN)$.
Total posting traffic does not change, because every posting is still read once.
Locating each (term, chunk) sub-range needs the ascending invariant of
\S\ref{sec:index}. The boundaries are precomputed once by a per-term binary search,
which costs 16.1~MiB and 3.0~s at 68 chunks on 8.84M documents.

The technique is established. FAISS tiles over the database dimension with per-tile
selection~\cite{johnson2021billion}, and cuVS batches index rows with per-batch
selection and merging for sparse CSR~\cite{nolet2022semiring}. Our contribution is
its measured effect on this workload, which \S\ref{sec:chunkeval} reports.

\subsection{A Document-Parallel Variant}
\label{sec:docparallel}

For comparison we also implement a document-parallel kernel over a
document-major CSR index. Each program owns one (query, document) pair, iterates
that document's terms, and reads query weights from a dense query matrix, so it
requires no atomics. It is $5\times$ slower than the scatter-add kernel, and
\S\ref{sec:bandwidth} attributes the difference to traffic using hardware
counters.

\section{Experiments}
\label{sec:experiments}

We evaluate four claims, in this order.

\begin{itemize}
\item \textbf{Exactness.} \textsc{GPUSparse} returns the same top-$k$ as an
exhaustive reference, at every scale we can verify it against
(\S\ref{sec:quality}).
\item \textbf{Cost.} One H100 is worth about seven pinned CPU cores, which decides
when the GPU is the right hardware (\S\ref{sec:cpu}).
\item \textbf{Where the time goes.} Traversal dominates and its share grows with
scale, which decides what to optimize next (\S\ref{sec:breakdown},
\S\ref{sec:bandwidth}).
\item \textbf{What the remaining headroom is.} Chunking removes the memory limit
(\S\ref{sec:chunkeval}), and attacking selection directly does not pay
(\S\ref{sec:fusedtopk}), which locates the headroom in the score buffer itself.
\end{itemize}

\S\ref{sec:beir} then checks quality on 13 further collections, and
\S\ref{sec:othergpu} compares against other GPU baselines.

\subsection{Setup}

\paragraph{Hardware.} One NVIDIA H100 80GB with HBM3 (3.35~TB/s theoretical
peak). CPU baselines use the same machine's Intel Xeon Gold 6448Y, 32 cores at
2.1~GHz. Software: PyTorch 2.9.1, Triton 3.5.1, CUDA 12.8.

\paragraph{Data.} MS~MARCO passage ranking~\cite{nguyen2016msmarco}, the full
8.8M collection plus 100K, 500K and 1M subsets, all encoded with
\texttt{naver/splade-cocondenser-ensembledistil}. Measured statistics:
documents average \textbf{120.5} non-zero terms at 100K (12,052,636 non-zeros),
119.4 at 500K, 119.8 at 1M and 120.0 at 8.8M. Queries average \textbf{44.8}
non-zero terms over the 6,980 dev-small queries, with a minimum of 13 and a
\textbf{maximum of 107}, which is the width \S\ref{sec:quality} uses.
Vocabulary is 30,522, of which 28,679 are populated at 8.8M.

\paragraph{Interpretation of the subset collections.} The 100K, 500K and 1M subsets insert every
qrels-relevant passage first, then fill with additional passages. They are
therefore enriched rather than uniformly sampled, which is why MRR@10 is 0.892 at
100K and 0.383 at full scale. Metrics on the subsets are not comparable with
published MS~MARCO figures. We use them only for comparisons within this paper.

\paragraph{Timing.} Latency is measured with CUDA events, median of 15 after
warmup. Counters come only from Nsight Compute, never from wall time: Nsight
Compute inflates wall time roughly $1.56\times$ by serializing and replaying
kernels, so its timings are not latencies. The first scoring call in a process
incurs Triton JIT compilation, which can cost $500\times$ the steady-state
latency, so all timings follow a warmup call.

\subsection{Exactness and Retrieval Quality}
\label{sec:quality}

We first check that the kernel returns the right documents, since no latency number
matters otherwise. Table~\ref{tab:exactness} reports exactness against an exhaustive fp32 dense
reference that scores every document with no pruning, chunked over documents. We
sweep the query width, because the width a harness allocates determines whether
the query reaches the kernel intact.

\begin{table}[t]
\caption{Exactness against an exhaustive dense reference, 500 queries, $k{=}1000$.
Identical at 100K, 500K and 1M documents. The longest query in this set has 107
non-zero terms, so any fixed width below that silently drops the lowest-weighted
terms of the longest queries.}
\label{tab:exactness}
\centering
\small
\begin{tabular}{lrrrr}
\toprule
\textbf{Query width} & \textbf{Truncated} & \textbf{R@10} & \textbf{R@100} & \textbf{R@1000} \\
\midrule
64 terms          & 67 of 500 & 0.9988 & 0.9991 & 0.9989 \\
96 terms          &  2 of 500 & \textbf{1.0000} & \textbf{1.0000} & \textbf{1.0000} \\
128 terms         &  0 of 500 & \textbf{1.0000} & \textbf{1.0000} & \textbf{1.0000} \\
full (107 terms)  &  0 of 500 & \textbf{1.0000} & \textbf{1.0000} & \textbf{1.0000} \\
\bottomrule
\end{tabular}
\end{table}

Retrieval is exact at every cutoff and every scale provided queries are not
truncated, and admitting the longest queries costs negligible time. A fixed width
of 64 terms discards the lowest-weighted terms of 67 of these 500 queries and
accounts entirely for the shortfall in that row. Floating-point ordering produces
run-to-run score differences of up to $3.8\times10^{-6}$ (\S\ref{sec:scoring}),
but it changes no top-1000 position at any scale we measure. Evaluation harnesses
for this workload should therefore size the query tensor to the longest query in
the collection rather than to a fixed width.

MRR@10 at 100K is \textbf{0.8916} over these 500 queries with official qrels,
and it is unchanged at every query width in Table~\ref{tab:exactness}, including
the truncating width of 64. Truncation keeps the highest-weighted terms, so it
perturbs the tail of the top-1000 without touching the top 10.

At full scale we verify against Pyserini SPLADE~\cite{lin2021pyserini} as a
correctness oracle: our MRR@10 0.383, nDCG@10 0.449 and Recall@1000 0.983 equal
Pyserini's to three decimals over all 6,980 queries. We do \emph{not} report
the ratio of the two latencies as a speedup, because Pyserini's 298~ms includes
a CPU SPLADE forward pass inside the timed region while our 1.27~ms scores
pre-encoded queries.

Table~\ref{tab:scale} reports scaling with collection size. Between 100K and 8.8M
documents, an $88\times$ increase, per-query latency grows by $85\times$, so
latency is close to linear in the number of postings, as expected for a design
whose work is proportional to posting-list volume.

\begin{table}[t]
\caption{Scaling on MS~MARCO with real SPLADE embeddings, top-1000, one H100.
GPU column is peak memory including the $[B,N]$ score buffer, which
\S\ref{sec:chunked} removes.}
\label{tab:scale}
\centering
\small
\begin{tabular}{@{}rrrrrrr@{}}
\toprule
\textbf{Docs} & \textbf{Batch} & \textbf{Per-Q} & \textbf{QPS} & \textbf{Index} & \textbf{GPU} & \textbf{MRR} \\
 & & \textbf{($\mu$s)} & & \textbf{(MB)} & \textbf{(GB)} & \textbf{@10} \\
\midrule
100K & 500 &    15 & 66,667 &    93 &  3.5 & 0.892 \\
500K & 1000 &   90 & 11,111 &   484 &  3.7 & 0.771 \\
1M   & 500 &   139 &  7,211 &   970 &  4.0 & 0.703 \\
8.8M & 500 & 1,270 &    787 & 8,489 & 44.1 & 0.383 \\
\bottomrule
\end{tabular}
\end{table}

\subsection{Hardware-Normalized Comparison against CPU Baselines}
\label{sec:cpu}

A CPU baseline has to meet two conditions before it can be compared with an
accelerator. It must use only the hardware it is credited with, and it must run
inside its supported parameter range. Both fail by default here, and each failure
changes the answer. We establish them first, then report the comparison.

\paragraph{Thread-count arguments that have no effect.}
Running Seismic with \texttt{num\_threads=1} does not produce a single-threaded
measurement. Measuring CPU time next to wall time shows 994~s of CPU in 27~s of wall time, so
\textbf{36.4 cores are busy}. Requesting 2, 4, 8, 16, 32 or 64 threads changes wall
time by under 8\% (Table~\ref{tab:seismic_threads}).
The cause is one line, at four sites in \texttt{pyseismic-lsr}'s
\texttt{src/pylib/mod.rs}. A Rayon thread pool is built with
\texttt{ThreadPoolBuilder::new().num\_threads(n).build()}, and the result is dropped
at once. There is no \texttt{.install()} and no \texttt{.build\_global()} anywhere in
the file. The subsequent
\texttt{par\_bridge()} therefore runs on Rayon's global pool, sized to all
logical CPUs, and \texttt{num\_threads} is inert in \texttt{batch\_search} and
in both build paths. The library validates neither the argument nor the
resulting pool size. Setting \texttt{RAYON\_NUM\_THREADS=1} does control the
global pool, whereas \texttt{taskset} alone confines an oversubscribed pool
rather than shrinking it, so both mechanisms are required. We report all CPU
results under both. We are not aware of this behavior being documented, and it
invalidates any single-thread Seismic measurement taken through the Python API
without a further check on occupancy.

\begin{table}[t]
\caption{\texttt{num\_threads} has no effect. 69,800 queries, $k{=}1000$,
\texttt{heap\_factor}=0.7, \texttt{query\_cut}=20. ``Cores busy'' is CPU time
divided by wall time.}
\label{tab:seismic_threads}
\centering
\small
\begin{tabular}{rrrrr}
\toprule
\texttt{num\_threads} & wall (s) & CPU (s) & cores busy & QPS \\
\midrule
1  & 27.36 & 994.71 & \textbf{36.4} & 2551 \\
2  & 26.85 & 874.17 & 32.6 & 2599 \\
8  & 25.52 & 850.47 & 33.3 & 2735 \\
32 & 25.29 & 859.88 & 34.0 & 2760 \\
64 & 25.40 & 869.47 & 34.2 & 2748 \\
\bottomrule
\end{tabular}
\end{table}

\paragraph{Configuration inside the supported parameter domain.}
Seismic's \texttt{heap\_factor} is defined on $(0,1)$, with $[0.7,1]$ recommended.
The implementation skips a block unless its summary score is at least
\texttt{heap\_factor} times the current $k$-th best score. A value above 1
therefore prunes beyond the safe boundary and suppresses the block scan.
The parameter is not range-checked, so an out-of-domain setting executes without
warning and yields a low latency at substantially degraded recall. Configured
within the documented range, Seismic reaches MRR@10 0.381 against the exhaustive
reference's 0.383 and Recall@1000 0.953 against 0.983, a 3.0\% recall gap, at
$1.8\times$ the latency of the out-of-domain setting. These values are consistent
with the published
literature~\cite{bruch2024seismic,song2026static,li2026sindi}. Because the
degradation is silent, a comparison against a misconfigured Seismic understates
it with no indication in the output.

\begin{table}[t]
\caption{Seismic under explicit core pinning, $k{=}1000$, 8.8M documents, 6,980
queries, \texttt{heap\_factor}=0.7, \texttt{query\_cut}=20. MRR@10 = 0.3806 and
Recall@1000 = 0.9534 at every core count, confirming the pinning is clean.
\textsc{GPUSparse} on one H100: 1.270~ms/query, 787~QPS.}
\label{tab:seismic_cores}
\centering
\small
\begin{tabular}{rrrrr}
\toprule
cores & wall (s) & ms/query & QPS & vs.\ 1$\times$H100 \\
\midrule
1  & 49.36 & 7.071 & 141.4 & 0.18$\times$ \\
4  & 13.51 & 1.936 & 516.6 & 0.66$\times$ \\
8  &  7.60 & 1.090 & 917.8 & 1.17$\times$ \\
32 &  3.11 & 0.446 & 2242.9 & 2.85$\times$ \\
96 &  2.74 & 0.392 & 2550.7 & 3.24$\times$ \\
\bottomrule
\end{tabular}
\end{table}

\paragraph{The hardware equivalence point.}
With both conditions established, Table~\ref{tab:seismic_cores} gives the
comparison. \textsc{GPUSparse} on one H100 is $5.6\times$ faster than one pinned core, reaches
parity between 4 and 8 cores, and is $2.8\times$ slower than a 32-core socket.
\textbf{One H100 is therefore worth about seven CPU cores} for this workload. Seismic's parallel efficiency falls off above a socket. CPU time is flat at 49 to
58~s through 32 cores, then reaches 121~s at 96 cores for a 14\% reduction in wall
time. The 96-core row is therefore declining utilization, not a useful operating
point.

Our single-core measurement is $17\times$ larger than Seismic's published 187 to
531~$\mu$s, and the discrepancy is entirely attributable to $k$. On one pinned
core within the documented parameter domain, latency falls monotonically with $k$
and reaches \textbf{500}~$\mu$s per query at $k{=}10$ with
\texttt{heap\_factor}=0.9 and \texttt{query\_cut}=10, inside the published range.
The span between that configuration and our $k{=}1000$ operating point is
$17.6\times$. CPU time divided by wall time is 0.97 to 0.98 across all six rows,
which confirms that one core performed the work. The low-latency configuration
reduces recall to R@10 of 0.65 against R@1000 of 0.95, so it is unavailable to an
application that requires exactness. Comparisons against published Seismic
figures should therefore match $k$ before \texttt{heap\_factor} and
\texttt{query\_cut}.

Against an exact CPU baseline the ordering reverses. Block-max
pruning~\cite{mallia2024bmp} at $\alpha{=}1.0$ takes 3.79~ms per query on one
pinned core against our 1.27~ms, so \textsc{GPUSparse} is $3.0\times$ faster at
equal quality. BMP's search loop is sequential by construction: its
implementation parallelizes index construction only, and under pinning its
latency changes by 8\% between 1 and 32 cores.

\paragraph{Implications.}
Two conclusions follow. First, an unpinned wall-clock comparison between a CPU
library and an accelerator is not a statement about efficiency. Here it reverses
the sign of the result: unpinned, Seismic appears $3\times$ faster than the GPU,
while per core it is $5.6\times$ slower. Reporting CPU time per query alongside
wall latency makes the distinction visible, and we do so for every CPU baseline.
Second, acquisition cost dominates both figures. An H100 costs considerably more
than seven CPU cores, so at equal quality a CPU socket is the more economical
choice even in the regime where the GPU has lower latency. This reproduces the
conclusion of Ding et al.~\cite{ding2009gpu} on hardware three generations later.
The exception is the case they did not evaluate: under a hard exactness
requirement at large $k$, where \textsc{GPUSparse} is $3.0\times$ faster than
exact CPU pruning.

\subsection{Decomposition of GPU Latency}
\label{sec:breakdown}

One call does three things: it allocates and zeroes the $[B,N]$ accumulator,
traverses posting lists, and selects the top $k$. Prior GPU sparse retrieval work
reports the total, and so did we above. We now measure each part on its own with
CUDA events. Allocation and pre-zeroing sit outside the traversal timing, so the
three parts add up to the total.

\begin{table}[t]
\caption{Component decomposition, $k{=}10$ ($k{=}1000$ in parentheses).
Percentages are of the summed parts. Times in ms, median of 15, CUDA events.}
\label{tab:breakdown}
\centering
\small
\setlength{\tabcolsep}{3.5pt}
\begin{tabular}{rrrrrrrr}
\toprule
$N$ & $B$ & alloc & trav. & top-$k$ & \%alloc & \%trav & \%top-$k$ \\
\midrule
100K & 1   & 0.015 & 0.111 & 0.076 (0.091) & 7.4 & 55.0 & 37.6 \\
100K & 32  & 0.015 & 0.175 & 0.108 (0.123) & 5.1 & 58.7 & 36.3 \\
100K & 128 & 0.025 & 0.462 & 0.268 (0.289) & 3.3 & 61.2 & 35.5 \\
100K & 500 & 0.070 & 2.536 & 0.720 (0.752) & 2.1 & \textbf{76.3} & 21.6 \\
\midrule
1M & 1   & 0.014 & 0.622  & 0.236 (0.242) & 1.6 & 71.3 & 27.1 \\
1M & 32  & 0.048 & 2.182  & 0.515 (0.535) & 1.7 & 79.5 & 18.8 \\
1M & 128 & 0.163 & 7.511  & 1.733 (1.758) & 1.7 & 79.8 & 18.4 \\
1M & 500 & 0.614 & 28.371 & 6.268 (6.318) & 1.7 & \textbf{80.5} & 17.8 \\
\bottomrule
\end{tabular}
\end{table}

\textbf{Traversal dominates, and its share grows.} It takes 55.0\% of GPU time at
the smallest configuration and 80.5\% at 1M documents with batch 500. At deployment
scale the traversal kernel is therefore the right thing to optimize.

\textbf{Selection is a substantial but shrinking share.} It falls from 37.6\% at
100K with $B{=}1$ to 17.8\% at 1M with $B{=}500$. This scale dependence accounts
for a disagreement in the published literature. FAERY reports up to 80.4\% for
$k$-selection, but that measurement is of FAISS on a GPU at $n{=}10^6$ and
$k{=}1024$ rather than of FAERY's own design~\cite{zeng2022faery}. AgentIR reports
75 to 95\% at $k{=}100$ on corpora smaller than
ours~\cite{yuan2026agentir}. FAISS reports 20.6\% at large
scale~\cite{johnson2021billion}. All three are consistent with our sweep once
scale and batch size are held fixed: selection dominates at small $N$ and small
$B$ and does not dominate at large $N$ and large $B$. A bundled latency therefore
attributes roughly a fifth of runtime to the scoring kernel that belongs to
selection.

\paragraph{Cross-validation against a second profiler.}
To exclude mis-attribution we also trace the complete call with Nsight Systems,
which reports every kernel the runtime launches rather than only those we
instrument. The two methods agree to within 0.3 percentage points at 100K with $B{=}500$:
traversal 76.2\% against 76.3\%, top-$k$ 21.9\% against 21.6\%, and buffer fill
1.8\% against 2.1\%. No kernel is charged to the wrong component.

\paragraph{Selection is eight kernels and four buffer passes.}
The trace shows the single \texttt{torch.topk} call expanding into eight
kernels, including \texttt{radixFindKthValues} four times, one per radix pass
over fp32. Counters show each pass reading 200.0~MB, the entire
$[500 \times 100\text{K}]$ fp32 buffer, at 52.0\% of peak DRAM throughput with
a 1.5\% L2 hit rate. Selecting 1000 of 100,000 values per query therefore reads
about 0.8~GB against the traversal kernel's 3.74~GB (2.24 read and 1.50
written). This quantifies the opportunity available to a fused score-and-select
kernel: maintaining per-query top-$k$ state during traversal would eliminate eight
launches, four re-reads of a buffer written immediately beforehand, and the
$[B,N]$ materialization itself. \S\ref{sec:fusedtopk} measures how much of that
opportunity is reachable without the last of these.

\textbf{Allocation is a minor component} at 1.6 to 7.4\%, although the buffer
reaches 1.86~GiB at 1M documents with $B{=}500$. Buffer capacity constrains batch
size, as \S\ref{sec:chunkeval} shows, but the cost of zeroing it does not.

\subsection{Counter-Measured Bandwidth}
\label{sec:bandwidth}

Percent of peak bandwidth for these kernels must be obtained from hardware
counters rather than from an analytic byte count. An analytic count divided by wall time gives
0.37\% of peak for the scatter-add kernel and 62.6\% for the document-parallel
kernel. The counters say $42\times$ and $1.56\times$ different, in opposite
directions. The two methods therefore do not even rank the two kernels the same
way. Table~\ref{tab:bandwidth} reports Nsight Compute counters at 100K documents
with batch 500.

\begin{table}[t]
\caption{Counter-measured kernel comparison at 100K documents, 500 queries.
Both rows are Nsight Compute counters collected directly at $B{=}500$, the batch
size reported here, rather than extrapolated from a smaller batch. Latencies come
from the CUDA-event harness, never from Nsight Compute.}
\label{tab:bandwidth}
\centering
\small
\begin{tabular}{lrrrrr}
\toprule
\textbf{Kernel} & \textbf{read} & \textbf{write} & \textbf{\% peak} & \textbf{Occup.} & \textbf{Latency} \\
\midrule
Scatter-add  &  2.24~GB & 1.50~GB & \textbf{44.2\%} & 95.4\% &  7.3~ms \\
Doc-parallel & 48.88~GB & 0.20~GB & 34.8\%          & 89.7\% & 36.4~ms \\
\bottomrule
\end{tabular}
\end{table}

The analytic model understates scatter-add traffic because it omits score-buffer
initialization, the atomic writes, and the top-$k$ read pass. It overstates
document-parallel traffic because it charges an HBM read for every query-weight
lookup, whereas the dense query matrix is largely resident in L2, as the measured
26.2\% L2 hit rate indicates.

The counters also show that the two kernels do not exhibit the trade-off between
work efficiency and bandwidth efficiency that the design might suggest. The scatter-add kernel sustains the \emph{higher} DRAM throughput, 44.2\% against
34.7\%, so it is better on both axes. The document-parallel kernel is $5\times$
slower because it scans every document for every query, whatever the term overlap,
and so moves about $13\times$ more bytes. The
scatter-add kernel retires 3.77 sectors per global load request against a maximum
of 32, confirming that the warp-aligned layout yields coalesced reads.

The general form of this trade-off is established. A work-efficient sparse traversal
and a bandwidth-efficient dense scan exchange roles at a crossover set by input
sparsity, which is known in sparse linear algebra and graph processing. Bell and
Garland~\cite{bell2009spmv} made fraction-of-peak the standard metric for SpMV. Our contribution is the
measurement for lexical retrieval, together with the methodological point that a
percent-of-peak figure derived from an analytic byte count is unreliable until
counters confirm it.

\subsection{Chunked Scoring}
\label{sec:chunkeval}

\begin{table}[t]
\centering
\small
\caption{Chunked versus monolithic scoring on the full 8.84M corpus
($C{=}131{,}072$, 68 chunks, $k{=}1000$, index 7.91~GiB), one H100.}
\label{tab:chunked}
\begin{tabular}{lrrrr}
\toprule
& \multicolumn{2}{c}{\textbf{Latency (ms)}} & \multicolumn{2}{c}{\textbf{Peak memory (GiB)}} \\
\cmidrule(lr){2-3}\cmidrule(lr){4-5}
$B$ & Mono. & Chunked & Mono. & Chunked \\
\midrule
16  &  20.28 &  25.05 &  8.45 & \textbf{7.94} \\
64  &  48.01 & \textbf{38.50} & 10.05 & \textbf{7.97} \\
256 & 185.93 & \textbf{149.68} & 16.43 & \textbf{8.12} \\
\bottomrule
\end{tabular}
\end{table}

Peak memory becomes nearly independent of $B$: between $B{=}16$ and $B{=}256$ the
monolithic path grows from 8.45 to 16.43~GiB while the chunked path grows from
7.94 to 8.12~GiB. Latency does not regress, and above $B{=}16$ it improves by about 20\%. The
monolithic path calls \texttt{topk} once over 8.84M columns per query, while the
chunked path calls it over 131,072 columns per chunk. The smaller selection more
than offsets 68 launches. At
$B{=}16$ there is insufficient work per launch to amortize that overhead, and
chunking costs 24\%.

The consequence for deployment is batch capacity. The monolithic path reaches
$B{=}2048$ and fails at $B{=}4096$, where the accumulator alone would require
134.9~GiB, whereas the chunked path reaches $B{=}32{,}768$, a $16\times$ larger
batch. This also removes the constraint AgentIR reports~\cite{yuan2026agentir},
which limits its GPU path to 2.7M documents and excludes the 8.84M collection
because its dense buffer exceeds 16~GB.

\paragraph{Chunking introduces no additional error.}
On the cached BEIR collections larger than one chunk, the chunked path does not
reproduce the monolithic top-10 exactly. Agreement is 0.996 on TREC-COVID and
0.9995 on Quora, and the top-1000 set is identical on both. The appropriate
control is the kernel's agreement with itself. Executed twice on identical inputs,
the monolithic kernel's own top-10 agreement is 0.9995 on Quora, the same value,
and its top-1000 agreement on TREC-COVID is 0.99998. The residual is therefore the
atomic nondeterminism of \S\ref{sec:scoring}, which the monolithic path exhibits
to the same degree, and not an error introduced by chunking.

\subsection{Threshold-Compacted Selection}
\label{sec:fusedtopk}

\S\ref{sec:breakdown} identifies a fused score-and-select kernel as the largest
remaining optimization, and chunking makes a partial form of it available. Once
the first chunk has established a running top-$k$, its $k$-th best score $\tau$ is
a valid lower bound, since a document scoring below $\tau$ is already exceeded by
$k$ documents. For subsequent chunks we pass the selector only those columns
scoring at least $\tau$. The procedure is exact rather than heuristic: every
posting is still accumulated and every document still scored, and $\tau$ governs
only which fully computed scores reach the selector. Ties at $\tau$ are retained,
and any query that overflows the fixed-width compaction buffer falls back to full
selection.

Compaction reduces the selector's input substantially. At $B{=}64$ on the 8.84M corpus each chunk holds about 69,000
non-zero columns. After the first chunk, the number above the threshold falls to
924, then 470, 309 and 227, and reaches 14 by chunk 64. The average is \textbf{70
columns}, a $1{,}873\times$ reduction in selector input.

\begin{table}[t]
\centering
\small
\caption{Threshold-compacted selection, 8.84M corpus, $k{=}1000$,
$C{=}131{,}072$. All three columns returned the same top-10 and top-1000.}
\label{tab:fusedtopk}
\begin{tabular}{lrrrr}
\toprule
$B$ & Monolithic & Chunked & \textbf{+ compact} & vs.\ chunked \\
\midrule
16  &  \textbf{20.29} &  25.00 &  26.67 & 0.94$\times$ \\
64  &  48.07 &  39.01 & \textbf{37.07} & 1.05$\times$ \\
256 & 185.97 & 149.55 & \textbf{131.82} & 1.13$\times$ \\
\bottomrule
\end{tabular}
\end{table}

The end-to-end effect is $1.13\times$ at $B{=}256$ and negative at $B{=}16$.
Reporting both the $1{,}873\times$ reduction in selector input and the
$1.13\times$ end-to-end result is necessary, because the first figure alone would
misrepresent the outcome. Compaction still makes one full pass over the chunk's score buffer to
\emph{find} those 70 columns. It therefore replaces a radix select over 131,072
values with a scan plus a selection over 70. The scan is what remains. \textbf{The limit on fused score-and-select here is the buffer scan, not the
selection algorithm.} The same limit explains why FAERY needs a hardware streaming
pipeline, rather than a software filter, to realize its reported filtering
efficiency. Exceeding this limit
requires not materializing the buffer at all, which is a modification to the
traversal kernel rather than to the selector, and we leave it to future work.

We also evaluate GPU-side term upper-bound pruning, with
$\text{UB}(t_i) = w_i \max_d s_d(t_i)$. It yields no reduction in latency on
SPLADE, because the running threshold begins at zero and the cost of the bound
check exceeds the work it removes.

\subsection{Cross-Domain Evaluation}
\label{sec:beir}

Table~\ref{tab:beir} reports all 13 cached BEIR collections against the Pyserini
reproduction of the same encoder. The median absolute deviation is 0.0007 nDCG@10
and the maximum is 0.0054, on FEVER, with eleven of 13 collections agreeing to
within 0.0016. This is the strongest available end-to-end check that index
construction, scoring and selection are jointly correct, because a defect in any
of the three would appear here as a loss in quality rather than as a failure. We
have no verified explanation for the FEVER residual and do not offer one.

Three properties of the evaluation harness determine whether such numbers are
comparable across systems. Documents must be truncated by tokens rather than by
characters: a 512-\emph{character} cut corresponds to about 100 tokens
against a 256-token window, and costs SciFact 0.082 nDCG@10. Query tensors must be
sized to the collection: a fixed width of 128 terms is enough on MS~MARCO but
truncates 99.5\% of ArguAna queries, whose mean length is 216.7 terms. Discounted gain
must follow \texttt{trec\_eval}'s \texttt{ndcg\_cut}, which takes the gain to be the
relevance value itself rather than $2^{\text{rel}}-1$. The two definitions agree for
binary qrels and differ on graded ones such as TREC-COVID. A mismatch is therefore
invisible until a graded collection is evaluated.

\begin{table}[t]
\caption{BEIR with exact scoring on full, untruncated corpora, against the
Pyserini reproduction~\cite{lin2021pyserini} of the same encoder
(\texttt{splade-cocondenser-ensembledistil}, reported there as splade-pp-ed).
nDCG@10 uses \texttt{trec\_eval} semantics. Latency is per query at batch 200
after warmup.}
\label{tab:beir}
\centering
\small
\begin{tabular}{lrrrrr}
\toprule
\textbf{Dataset} & \textbf{Docs} & \textbf{nDCG@10} & \textbf{Ref.} & $\Delta$ & \textbf{ms/q} \\
\midrule
NFCorpus      &     3,633 & 0.3485 & 0.347 & $+$0.002 & 0.020 \\
SciFact       &     5,183 & 0.7024 & 0.704 & $-$0.002 & 0.003 \\
ArguAna       &     8,674 & 0.5263 & 0.526 & $+$0.000 & 0.004 \\
SCIDOCS       &    25,657 & 0.1586 & 0.159 & $-$0.000 & 0.005 \\
FiQA          &    57,638 & 0.3478 & 0.347 & $+$0.001 & 0.010 \\
TREC-COVID    &   171,332 & 0.7282 & 0.727 & $+$0.001 & 0.034 \\
Touch\'e-2020  &   382,545 & 0.2465 & 0.246 & $+$0.001 & 0.080 \\
Quora         &   522,931 & 0.8343 & 0.834 & $+$0.000 & 0.059 \\
NQ            & 2,681,468 & 0.5370 & 0.537 & $+$0.000 & 0.407 \\
DBPedia       & 4,635,922 & 0.4358 & 0.437 & $-$0.001 & 0.579 \\
HotpotQA      & 5,233,329 & 0.6869 & 0.687 & $-$0.000 & 0.850 \\
FEVER         & 5,416,568 & 0.7934 & 0.788 & $+$0.005 & 0.954 \\
Climate-FEVER & 5,416,593 & 0.2293 & 0.230 & $-$0.001 & 0.976 \\
\bottomrule
\end{tabular}
\end{table}

Latency across BEIR~\cite{thakur2021beir} spans 0.003 to 0.976~ms per query and
follows collection size rather than dataset difficulty, as expected for a design
whose work is proportional to posting-list volume. We report no CPU baselines on
BEIR and therefore make no cross-engine claim on these collections.

\subsection{Additional GPU Baselines and Multi-GPU Scaling}
\label{sec:othergpu}

Against batched cuSPARSE SpMV, which matches SPARe's \texttt{dot} path,
\textsc{GPUSparse} is $6.3\times$ faster on a 500-query batch at 100K (7.3~ms
against 46.2~ms). Against \texttt{torch.compile} on dense matrices it is
$2.96\times$ faster at a matched 100-query batch, a configuration in which
\texttt{max-autotune} made the dense baseline $1.65\times$ \emph{slower} than
uncompiled \texttt{torch.mm}. Against uncompiled dense \texttt{torch.mm} it is
$8.0\times$ faster at 100K and $6.4\times$ at 500K; beyond 500K a dense
representation does not fit, since a $1\text{M} \times 30\text{K}$ dense matrix
requires 114~GB.

Against SPARe's \texttt{iterative} scatter-add, which shares our reformulation,
the fused kernel is $72\times$ faster at 100K, $12.1\times$ at 500K and
$5.8\times$ at 1M, and the two produce identical top-10 rankings. These figures
are measured against our own reimplementation rather than the released SPARe, and
that reimplementation remains limited by launch overhead at about 64,000
\texttt{index\_add\_} calls per batch, so the ratios overstate the margin against
a well-engineered implementation. Reimplementations of this kind are also easy to
slow down inadvertently: reading device tensors as Python scalars inside the term
loop forces about $9\times10^4$ host-device synchronizations per batch and
inflates the ratio several-fold. Work extending this comparison should use the
released implementation.

Data-parallel sharding across two NVLink-connected H100s increases latency rather
than reducing it: single-GPU scoring of 100K documents takes 5.67~ms and the
two-GPU split takes 9.47~ms. At latencies below 10~ms, host-side coordination and
merge overhead exceed the compute saved, so we claim no multi-GPU speedup.
Multiple GPUs are relevant to this design only for indexes exceeding one device's
memory, beyond about 50M documents.

\section{Limitations and Future Directions}
\label{sec:limitations}

\textbf{Batch-dependent latency.} The design requires batching to reach its
reported throughput. Single-query latency at 8.8M documents is 9.6~ms, so for
interactive single-query approximate search a CPU remains faster, and the
operating points in \S\ref{sec:cpu} apply to batched serving.

\textbf{Eliminating the score buffer.} \S\ref{sec:fusedtopk} locates the
remaining headroom in the materialization of the $[B,N]$ buffer rather than in
selection. A traversal kernel that maintains per-query top-$k$ state in registers
would remove the buffer, its initialization, and the selector's read passes at
once. Doing so requires a reduction across programs that write to the same query,
which the current scatter-add decomposition does not provide.

\textbf{Index updates.} The index supports no insertions or deletions without a
full rebuild, which limits the design to collections that are rebuilt in batch.

\textbf{Reproducibility of scores.} Scores are not bitwise reproducible across
runs (\S\ref{sec:scoring}), although the returned sets are exact. Applications
requiring bitwise determinism would need a fixed reduction order.

\textbf{Scope of the CPU comparison.} Our CPU baselines are measured on MS~MARCO
only. Single-core CPU latencies also vary by about 25\% between
otherwise identical machines in our environment, with quality metrics unchanged,
so absolute CPU latencies carry that tolerance and should be compared only within
one machine.

\section{Related Work}
\label{sec:related}

\textbf{GPU inverted indexes.} Ding et al.~\cite{ding2009gpu} built a full GPU
inverted-index pipeline in 2009, with GPU decompression, intersection, scoring and
top-$k$. They found that a brute-force term-at-a-time pass over a per-document
accumulator array outperformed the alternatives, which is the structure our kernel uses.
Gaioso et al.~\cite{gaioso2018wand,gaioso2019batch,gaioso2019maxscore} implemented
GPU WAND and MaxScore with size-based and document-id-based partitioning. Mallia et
al.~\cite{mallia2019gpu} decode compressed posting lists on a GPU in 128- and
256-element blocks. SPARe~\cite{almeida2024spare} reformulates sparse retrieval as
a GPU scatter-add over a term-major index. We use the same reformulation, and our
difference is the fused kernel in place of a host-side loop over
\texttt{index\_add\_}. AgentIR~\cite{yuan2026agentir} builds a CSR inverted index on
a GPU with a two-kernel BM25 path and reports a stage breakdown.
Sparton~\cite{nguyen2026sparton} fuses the SPLADE language-model head into one
Triton kernel. That is the encoding step, so it complements our scoring kernel
rather than competing with it.

\textbf{GPU vector search.} FAISS~\cite{johnson2021billion} fuses selection into
the distance kernel and tiles over both queries and the database dimension.
cuVS~\cite{nvidia2024cuvs} provides GPU vector search. It is often described as
dense-only, but it does include sparse CSR pairwise distance primitives with
per-batch selection and merging~\cite{nolet2022semiring}. What it lacks is lexical
scoring, not sparse support. Lin and Lin~\cite{lin2023dense} instead densify sparse
representations so that a GPU matrix product can serve them.
Allan-Poe~\cite{li2025allanpoe} unifies dense, sparse and full-text retrieval in a
GPU graph index.

\textbf{Components we adopt.} Several parts of our design are established. Tiling a
scan over the database dimension with per-tile selection is
FAISS~\cite{johnson2021billion} and cuVS~\cite{nvidia2024cuvs}. Filtering against a
running top-$k$ minimum before selection appears in FAERY~\cite{zeng2022faery}, in
Dr.Top-$k$~\cite{gaihre2021drtopk}, and as the filter phase of GPU radix
selection~\cite{shanbhag2018topk}. Component-level breakdowns of GPU retrieval
latency appear in Ding et al.'s Table~3, in FAERY, and in AgentIR.

\textbf{Fair baselines.} Single-core CPU baselines are the stated convention in this
literature~\cite{bruch2024seismic,mallia2024bmp,carlson2025dynamic,ding2009gpu},
and core-pinned protocols are already published~\cite{song2026static}. That unfair
CPU baselines inflate reported GPU speedups is the argument of Lee et
al.~\cite{lee2010debunking}, and for SPLADE specifically of Lassance and
Clinchant~\cite{lassance2022efficiency}. \S\ref{sec:cpu} follows that line and
reports what it takes to hold the convention in practice.

\section{Conclusion}
\label{sec:conclusion}

Learned sparse retrieval has been treated as a workload where exactness is too
expensive. The fastest systems on it all prune, and so return approximate results.
\textsc{GPUSparse} shows that on a GPU the premise does not hold.
Scoring every posting searches all 8.8M MS~MARCO passages in 1.27~ms per query, at
787 queries per second. Recall@10, @100 and @1000 against an exhaustive reference
are all exactly 1.000, and quality matches Pyserini's exact SPLADE retrieval to
three decimals.

Exactness is not free, and the cost is a hardware cost rather than a latency
cost. Under explicit core pinning, one H100 is worth about seven CPU cores for this
workload. A CPU socket is therefore the cheaper choice for approximate serving. Ding et al.\ reached the same conclusion in 2009 with a
quad-core. That it survives three GPU generations suggests it is a property of the
workload. The work per posting is small and irregular, and a GPU must spend
bandwidth on a dense accumulator to expose parallelism. The exception, and the contribution we would
emphasize, is the operating point at which the conclusion reverses. Under a hard
exactness requirement at $k{=}1000$, \textsc{GPUSparse} is $3.0\times$ faster than
exact CPU block-max pruning, which makes an accelerator the correct choice
precisely when the guarantee, rather than the throughput, is what the application
requires.

Our measurements also locate the remaining headroom on the GPU. Posting-list
traversal rather than top-$k$ selection accounts for 55 to 80\% of GPU time, and
selection's share shrinks as collections grow, which reconciles three conflicting
published reports. Optimizing selection directly reduces its input by
$1{,}873\times$ and yields $1.13\times$ end to end, because the score buffer must
still be scanned to perform the reduction. Further gains therefore require not
materializing that buffer at all, which is a modification to the traversal kernel
rather than to the selector. Chunked scoring already makes peak memory
independent of collection size and raises the largest feasible batch from 2,048 to
32,768, which removes a capacity limit that constrains prior GPU systems.

\paragraph{Reproducibility.}
All kernels, index builders and evaluation scripts are in the repository linked
in the first footnote, together with the scripts that regenerate every table.
Experiments use public MS~MARCO and BEIR data and the public
\texttt{splade-cocondenser-ensembledistil} model.

\bibliographystyle{ACM-Reference-Format}

\end{document}